\documentclass[twocolumn,twoside,slac_two]{revtex4}
\usepackage{graphicx}
\usepackage{fancyhdr}
\begin{document}

\title{Pinpointing the TeV gamma-ray emission region in M87 using TeV and 43 GHz radio monitoring}

\author{R. M. Wagner}
\affiliation{Max-Planck-Institut f\"ur Physik, D-80805 Munich, Germany}
\affiliation{Excellence Cluster ``Universe'', D-85748 Garching, Germany}
\author{M. Beilicke}
\affiliation{Washington University, St. Louis, MO 63130, USA}
\author{F. Davies}
\affiliation{National Radio Astronomy Observatory, Socorro, NM 87801, USA}
\author{P. Hardee}
\affiliation{University of Alabama, Tuscaloosa, AL 35487, USA}
\author{H. Krawczynski}
\affiliation{Washington University, St. Louis, MO 63130, USA}
\author{D. Mazin}
\affiliation{Institut de Fisica d'Altes Energies, E-08193 Bellaterra (Barcelona), Spain}
\author{R. C. Walker}
\affiliation{National Radio Astronomy Observatory, Socorro, NM 87801, USA}
\author{M. Raue}
\affiliation{Max-Planck-Institut f\"ur Kernphysik, D-69117 Heidelberg, Germany}
\author{S. Wagner}
\affiliation{Landessternwarte Heidelberg, D-69117 Heidelberg, Germany}
\author{C. Ly}
\affiliation{University of California, Los Angeles, CA 90095-1547, USA}
\author{W. Junor}
\affiliation{Los Alamos National Laboratory, Los Alamos, NM 87545, USA}
\author{for the MAGIC, VERITAS, H.E.S.S. collaborations}

\begin{abstract}
The TeV radio galaxy M87 is the first radio galaxy detected in the TeV regime.
The structure of its jet, which is not pointing towards our line of sight, is
spatially resolved in X-ray (by Chandra), optical and radio observations. In
2008, the three main Atmospheric Cherenkov Telescope observatories VERITAS,
MAGIC and H.E.S.S. coordinated their observations in a joint campaign from
January to May with a total observation time of approx. 120 hours. In February,
strong and rapid day-scale TeV flares were detected. VLBA monitoring
observations during the same period showed that the 43 GHz radio flux density
of the unresolved core began to rise at the time of the TeV flares and
eventually reached levels above any previously seen with VLBI. New jet
components appeared during the flare. The localization accuracy of the TeV
instruments of many arcseconds, even for strong sources, is inadequate to
constrain the origin of the emission in the inner jets of AGNs. For M87, with a
6 billion solar mass black hole and a distance of 16.7 Mpc, the VLBA resolution
instead corresponds to 30 by 60 Schwarzschild radii. This is starting to
resolve the jet collimation region. The temporal coincidence of the TeV and
radio flares indicates that they are related and provides the first direct
evidence that the TeV radiation from this source is produced within a few tens
of $R_S$ of the radio core, thought to be coincident to within the VLBA resolution
with the black hole.

\end{abstract}

\maketitle

\thispagestyle{fancy}

\section{M\,87 AS A UNIQUE LABORATORY FOR BLAZAR ASTROPHYSICS}

Active galactic nuclei (AGN) are extreme extragalactic objects showing
core-dominated emission (broadband continuum ranging from radio to X-ray
energies) and strong variability on different timescales. A supermassive black
hole (in the center of the AGN) surrounded by an accretion disk is believed to
power the relativistic plasma outflows (jets) which are found in many AGN.
More than 32 AGN have been found to emit VHE $\gamma$-rays (E$>$100\,GeV).\footnote{See, e.g., {\tt http://www.mpp.mpg.de/$\sim$rwagner/sources/}}  The
size of the VHE $\gamma$-ray emission region can generally be constrained by
the time scale of the observed flux variability \cite{Mrk421_Burst,Aha06} but
its location remains unknown. 

The giant radio galaxy M\,87 is located at a distance of $16.7 \, \mathrm{Mpc}$
(50 million light years) in the Virgo cluster of galaxies \cite{Mac99}. The
angle between the plasma jet in M\,87 and the line of sight is estimated to lie
between $20^{\circ}-40^{\circ}$ \cite{Bir95,Bir99}.  With its proximity, its
bright and well resolved jet, and its very massive black hole with $(6.0\pm0.5) \times
10^{9} \, \rm{M}_{\odot}$ \cite{BH_M87}, M\,87 provides an excellent
opportunity to study the inner structures of the jet, which are expected to
scale with the gravitational radius of the black hole. Substructures of the jet
are resolved in the X-ray, optical and radio wavebands \cite{ChandraSpecM87}
and high-frequency radio very long baseline interferometry (VLBI) observations
with sub-milliarcsecond (mas) resolution are starting to probe the collimation
region of the jet \cite{JetFormation}. This makes M\,87 a unique laboratory in
which to study relativistic jet physics in connection with the mechanisms of
VHE $\gamma$-ray emission in AGN.

VLBI observations of the M\,87 inner jet show a well resolved, edge-brightened
structure extending to within $0.5 \, \rm{mas}$ ($0.04 \, \rm{pc}$ or $70$
Schwarzschild radii $R_S$) of the core. Closer to the core, the jet has
a wide opening angle suggesting that this is the collimation region
\cite{JetFormation}.  Along the jet, monitoring observations show both
near-stationary components \cite{InnerJet} (pc-scale)  and features that move
at apparent superluminal speeds \cite{HST_Superluminal,Jet_And_TeV} (100
pc-scale). The presence of superluminal motions and the strong asymmetry of the
jet brightness indicate that the jet flow is relativistic. The near-stationary
components could be related to shocks or instabilities that can be either
stationary or move more slowly than the bulk flow.

\section{TEN YEARS OF VHE GAMMA-RAY OBSERVATIONS OF M\,87}

Currently, there are more than 30 extragalactic objects~-- all belonging to the
class of AGN~-- that have been established as VHE $\gamma$-ray emitters by
ground-based imaging atmospheric Cherenkov telescopes (IACTs), such as
 H.E.S.S. \cite{Hin04}, MAGIC \cite{Lor04} and VERITAS \cite{Acc08a}.
So far, all of them except the radio galaxies
M\,87, Centaurus A \cite{CenA}, and possibly 3C~66B \cite{3C66B}, as well as the starburst galaxies M82 \cite{m82} and NGC 253 \cite{ngc253}, belong to the
class of blazars (exhibiting a plasma jet pointing closely to our line of
sight). 

\begin{figure}
\includegraphics[width=\linewidth]{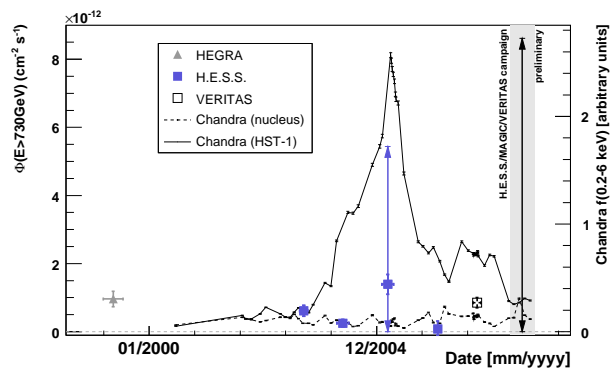}
\caption{The yearly averaged VHE $\gamma$-ray light curve $\Phi(E>730 \,
\rm{GeV})$ of M\,87, covering the 10-year period from 1998-2008. The data
points are taken from \cite{Aha03,Aha06,Acc08b,Alb08}. Vertical arrows indicate
the measured flux ranges for the given period (if variable emission was found).
Strong variabilities ($<$ 2 days flux doubling times) were measured in
$\gamma$-rays in 2005 and 2008.  The Chandra X-ray light curves of the nucleus
and HST-1 are also shown \cite{Har06}.}
\label{fig:LC_All}
\end{figure}

A first indication of VHE $\gamma$-ray emission ($> 4 \, \sigma$) from the
direction of M\,87 in 1998/9 was reported by HEGRA \cite{Aha03}. The VHE
$\gamma$-ray emission was confirmed by H.E.S.S. \cite{Aha06}, establishing
M\,87 as the first non-blazar extragalactic VHE $\gamma$-ray source. The
reported day-scale variability strongly constrains the size of the $\gamma$-ray
emission region.  VERITAS detected M\,87 in 2007 \cite{Acc08b} but with no
variability.  Recently, the short-term variability in M87 was confirmed by
MAGIC in a strong VHE $\gamma$-ray outburst \cite{Alb08}. The yearly
averaged VHE $\gamma$-ray light curve of M\,87 for the past 10 years is shown
in Fig.~\ref{fig:LC_All}.  The measured flux variability rules out large-scale
emission from dark matter annihilation \cite{Bal99}, or cosmic-ray interactions
\cite{Pfr03}. Leptonic \cite{Geo05,Len08} and hadronic \cite{Rei04} jet
emission models have been proposed to explain the TeV emission.  The location
of the VHE $\gamma$-ray emission is still unknown, but the nucleus
\cite{Ner07}, the inner jet \cite{Geo05,Rei04,Len08,Tav08} or larger structures
in the jet \cite{Che07} have been proposed as possible sites. The 2005 VHE
$\gamma$-ray flare (H.E.S.S.) was detected during an exceptional, several years
lasting X-ray outburst of the innermost knot in the jet ``HST-1'' \cite{Har06},
whereas the recent VHE $\gamma$-ray flaring activity (reported here) occurred
during an X-ray low state of HST-1 (see Fig.~\ref{fig:LC_All}). In this paper
we report on a joint VHE observation campaign of M\,87 performed by H.E.S.S.,
MAGIC and VERITAS in 2008.

\begin{figure}
\centering
\includegraphics[width=\linewidth]{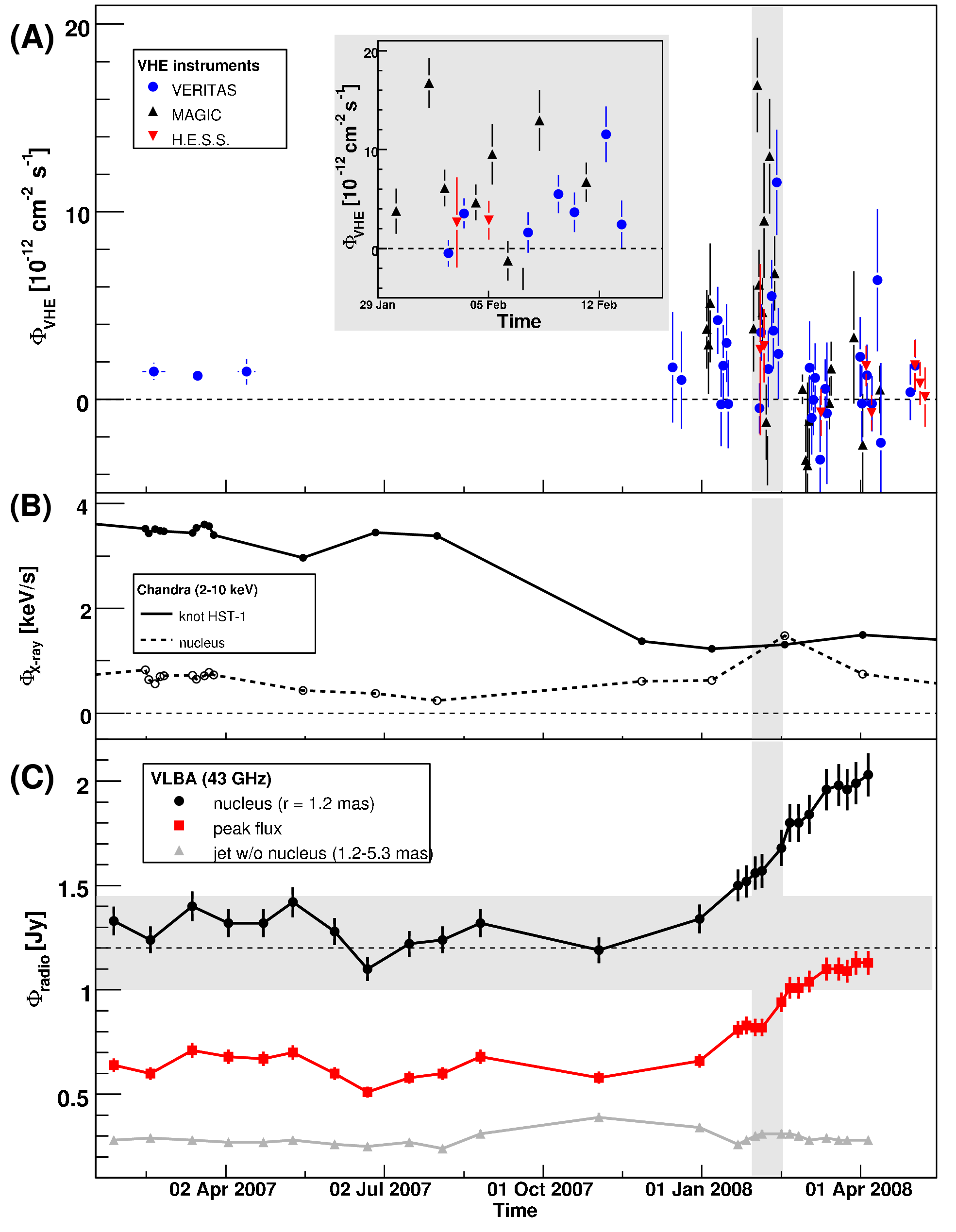}
\caption{{\it (A)}: the night-by-night averaged VHE $\gamma$-ray light
curve $\Phi(E>350 \, \rm{GeV})$ of M\,87, covering the 2008 joint campaign.
Strong variability resulted in a detection of at least two flares (see the
inlay for a enlarged view of the relevant period).
{\it (B)}: Corresponding \textit{Chandra} measurements
of the core and the HST-1 knot of M87.
{\it (C,D)}:
Flux densities from the 43 GHz VLBA observations for the nucleus, the peak flux
(VLBA resolution element), and the flux integrated along the jet.  The shaded
horizontal area indicates the range of fluxes from the nucleus before the 2008
flare. While the flux of the outer regions of the jet does not change
significantly, most of the flux increase results from the region around the
nucleus.
}
\label{fig:LC_2008}
\end{figure}

\section{THE 2008 CAMPAIGN ON M87}

\subsection{Joint H.E.S.S./MAGIC/VERITAS VHE observations}

IACTs measure very high energy ($E > 100$~GeV) $\gamma$-rays.  The angular
resolution of $\sim$0.1$^\circ$ of IACTs does not allow to resolve the M\,87
jet, but the time scale of the VHE flux variability constrains the size of the
emission region, while flux correlations with observations at other wavelengths
may enable conclusions on the location of the VHE $\gamma$-ray source.  The
current generation of instruments requires less than $10 \, \rm{h}$ for the
detection of a faint source with a flux level of a few percent of the Crab
nebula flux.  For a variable VHE $\gamma$-ray source like M\,87, a joint
observation strategy as well as combining the results from several IACT
experiments (and observations at other wavelengths) can substantially improve
the scientific output (e.g.  \cite{Maz05}). Coordinated observations with the
VHE instruments H.E.S.S., MAGIC, and VERITAS result in:
\begin{itemize} 
\item an extended energy range by combining data sets
taken under different zenith angles \item an extended visibility during one
night, because the visibility of any given celestial object depends on the
longitude of the experimental site \item an improved overall exposure and
homogeneous coverage of the source \item alerts and direct follow-up
observations in case of high flux states.
\end{itemize}

\begin{figure*}
\includegraphics[width=0.45\textwidth]{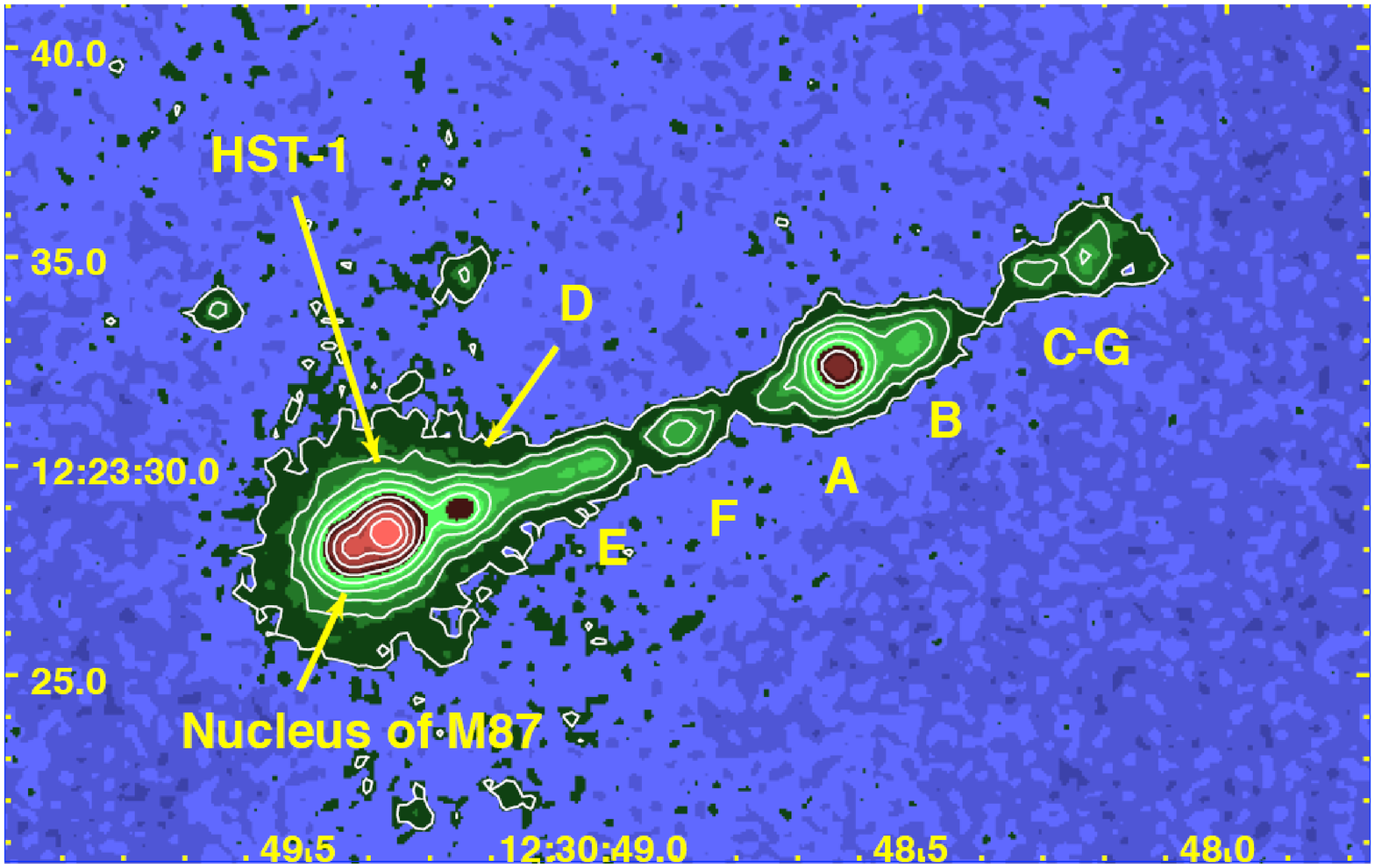}
\includegraphics[width=0.54\textwidth]{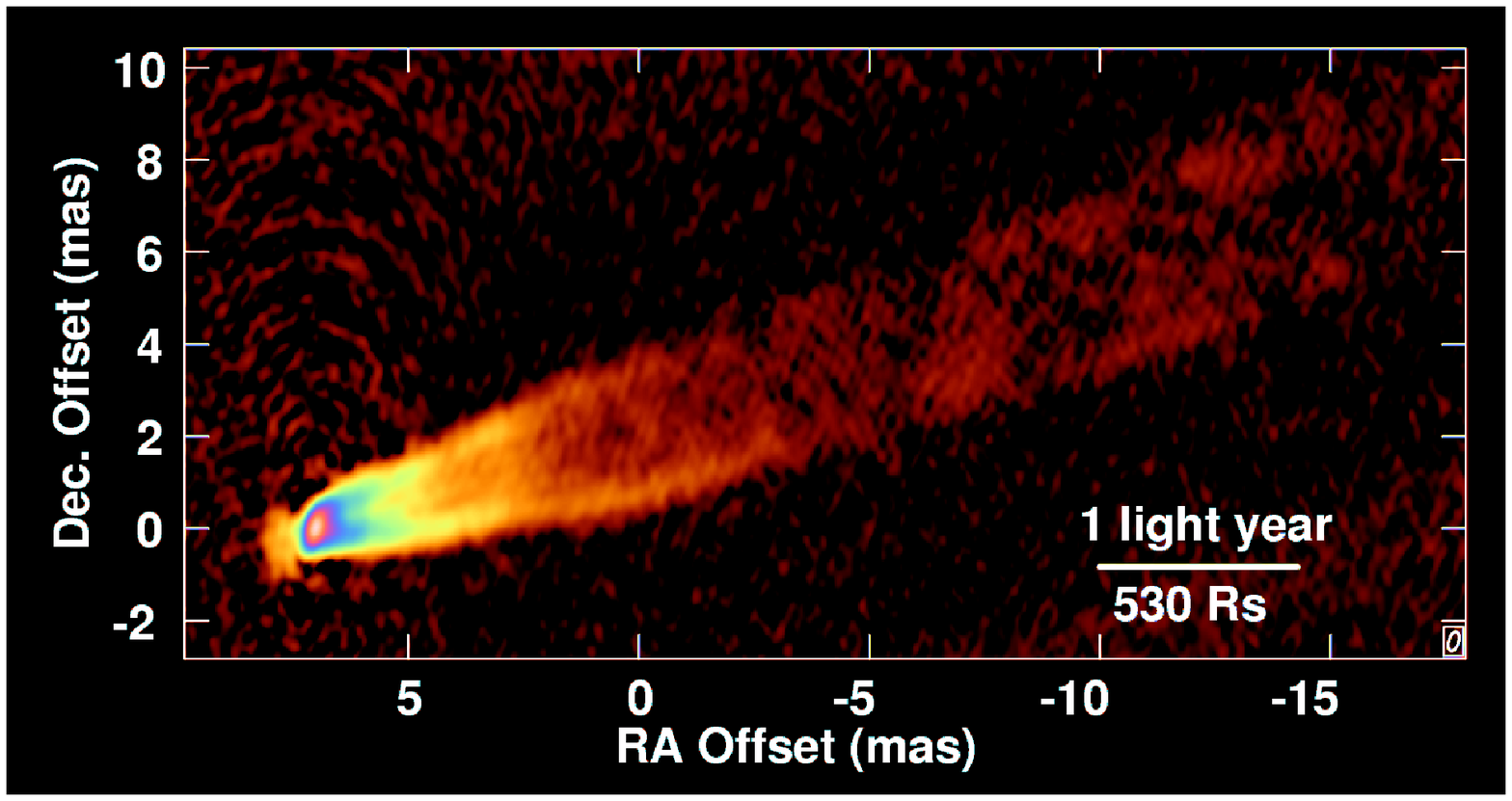}
\caption{Image of the M\,87 jet with high resolution instruments: Large-scale
jet in X-ray obtained with Chandra (left). Inner jet in radio (43\,GHz) obtained with VLBA (right).
\label{fig:X}}
\end{figure*}

Among other AGN, the radio galaxy M\,87 is part of a multi-collaboration AGN
trigger agreement between H.E.S.S., MAGIC and VERITAS. In order to achieve a
best possible VHE coverage (especially during Chandra X-ray observations) a
closer cooperation was conducted for the 2008 observations of M\,87. The
collaborations agreed to have a detailed exchange/synchronization of their
M\,87 observation schedules.  Further on, information about the status of the
observations (i.e. loss of observation time due to bad weather conditions,
etc.) was exchanged on a regular basis between the shift crews and the
observation coordinators. 

M\,87 was observed by the three experiments for a total of $>120 \, \rm{h}$ in
2008 ($\sim$95~h after quality selection). The amount of data resulted in an
unprecedentedly good coverage with $>$50 nights between January and May 2008. 

\subsection{Chandra X-ray observations}
Chandra monitoring of M\,87 began in 2002 and continues to date. The angular
resolution ($\approx$\,0.8$^{\prime\prime}$) allows resolving the large-scale
jet structure, and in particular to distinguish emission from the core and the
innermost knot `HST-1' (left panel in Fig.~\ref{fig:X}). During an observing
season, M\,87 is observed every $\sim$6~weeks. That sampling allows detection
of 1.5\,month-scale variability.  The most remarkable discovery of the
monitoring campaign so far has been the giant flare of HST-1 \cite{Har06},
which reached its maximum intensity in 2005 (Fig.~\ref{fig:LC_All}) when the TeV emission was
detected in flaring state for the first time, suggesting HST-1 as the possible
origin of the VHE emission \cite{Che07}.  Simultaneously, a huge optical flare
was detected by the Hubble Space telescope \cite{HST_M87}.  Additional
observations were taken in 2007 on shorter intervals to investigate short-time
variability and possible correlation with VHE emission, which was unfortunately
in a quiet state at that time.

\begin{figure}
\includegraphics[width=\linewidth]{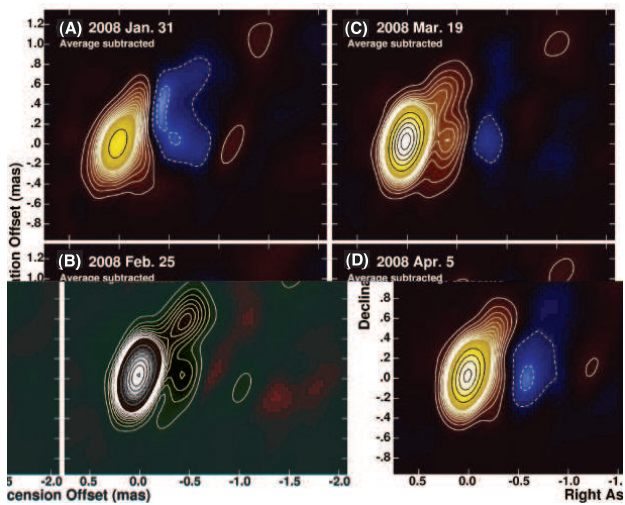}
\caption{
VLBA images of M 87 at 43 GHz. 
(A)-(D): Sequence of difference images for observations during the period of the radio 
flare.  These images have had an average image subtracted in order to show the effects of the flare.  
The average was made from eleven observations in 2007, outside the period affected by the flare.
The contours are linear with 10 (white) at intervals
of $7 \, \rm{mJy}$ per beam followed by the rest (black) at intervals of
$70 \, \rm{mJy}$ per beam; negative contours are indicated by dashed lines. 
The sequence shows the significant rise in the core flux density and the appearance of 
enhanced emission along the inner jet. 
\label{fig:radio}}
\end{figure}

\subsection{Radio: VLBA}
 Throughout 2007, M\,87 was observed at 43\,GHz with the VLBA on a regular
basis roughly every three weeks \cite{Supp_VLBA_MovieM87}.  In January 2008,
the campaign was intensified to one observation every 5 days.  The resolution
of the observations is rather high with 0.21$\times$0.43 milliarcseconds or
15$\times$30 Schwarzschild diameters of M\,87.  The aim of this ``movie project''
was to study morphological changes of the plasma jet with time. Preliminary
analysis of the first 7 months showed a fast evolving structure, somewhat
reminiscent of a smoke plume, with apparent velocities of about twice the speed
of light. These motions were faster than expected so the movie project was
extended from January to April 2008 with a sampling interval of 5 days.

The 43 Ghz radio flux density from the unresolved core rose by 0.3 Jy (36\%) while the integrated flux density from within 1.2 mas of the core rose by 0.57 Jy (32\%). 
beginning at the time of the VHE flare and extending over at least the
following two months until the VLBA monitoring project ended (Fig.~\ref{fig:LC_2008}).
Beyond 1.2 mas, there was no change.
The initial radio flux density increase was located in the unresolved core. The region around the core
brightened as the flare progressed (Fig.~\ref{fig:radio}), suggesting that new components
were emerging from the core. At the end of the observations, the brightened
region extended about 0.77 mas from the peak of the core implying an average
apparent velocity of $1.1 c$, well under the
approximately $2.3 c$ seen just beyond that distance in the first half of 2007.
The position of the M 87 radio peak did not move by more
than $\approx 12 R_S$ during the flare, indicating that
the peak emission corresponds to the nucleus of M 87.

\subsection{Results}

In January 2008, the VHE $\gamma$-ray flux was measured at a slightly higher
level than in 2007.  MAGIC detected a strong flaring activity in February 2008
\cite{Alb08}, which led to immediate intensified observations by all
three VHE experiments.  VERITAS detected another flare about one week after the
MAGIC trigger. The joint 2008 VHE $\gamma$-ray light curve clearly confirms the
short-term variability reported by H.E.S.S. in 2005 \cite{Aha06}. During the
2008 VHE flaring activity, MAGIC observed flux variability above $350 \,
\rm{GeV}$ on time scales as short as 1~day (at a significance level of
$5.6$~standard deviations). At lower energies ($150 \, \rm{GeV}$ to $350 \,
\rm{GeV}$) the emission was found to be compatible with a constant level
\cite{Alb08}.  From March to May, M\,87 was back in a quiet state.

In 2008, the X-ray and VHE $\gamma$-ray data suggested a different picture
compared to the 2005 flare (Fig.~\ref{fig:LC_All}): HST-1 was in a low state,
with the flux being comparable with the X-ray flux from the core. The core,
however, showed an increased X-ray flux state in February 2008, reaching a
highest flux ever measured with Chandra just few days after the VHE flaring
activity.

VLBA measured a continuously increasing radio flux from the region of the
nucleus ($r$=1.2 mas) during the 2008 campaign, whereas in 2007 the flux was
found to be rather stable.  Individual snapshots of the inner region of the jet
are shown in Fig.~\ref{fig:radio}.  The observed radio flux densities reached
at the end of the 2008 observations, roughly 2 months after the VHE flare
occurred, are larger than seen in any previous VLBI observations of M\,87 at
this frequency, including during the preceding 12 months of intensive
monitoring, in 6 observations in 2006 and in individual observations in 1999,
2000, 2001, 2002, and 2004 \cite{Supp_VLBA_M87}.  This suggests that radio
flares of the observed magnitude are uncommon. 

Given the rare occurrence of VHE, radio and X-ray flares at very
similar times as compared to their characteristic time scales, we conclude that
the events are likely connected. This conclusion is supported by the joint
modeling of the VHE and radio light curves (Fig. 5). The VLBI structure
of the flare, along with the timing of the VHE activity, is strong evidence
that the VHE emission occurred in a region small compared to the VLBA
resolution. The observed pattern can be explained by an event in the central
region causing the VHE flare. The effect of synchrotron~­self absorption causes
a delay of the observed peak radio emission since the region is not transparent
at radio energies at the beginning. This will lead to a smoothed and delayed
shape of the radio light curve. 

M 87 is the first radio galaxy that shows evidence for a connection between
simultaneous, and well sampled, radio and VHE $\gamma$-ray flux variations (which are
separated in photon frequency by 16 orders of magnitude), opening a new avenue
for the study of the AGN accretion and jet formation. General relativistic
magneto­ hydrodynamic simulations indicate that jets might be launched and
collimated magnetically over large spatial scales of $\approx 1000 R_S$, e.g. \cite{McK06}.
In the radio galaxy M 87, the VHE emission seems to originate much closer to the central region. Either it originates directly in the black hole magnetosphere ­ assuming that the radio core is coincident with the black hole ­ or it originates in a jet which accelerates on spatial scales smaller than $\approx 100 R_S$.

The correlation study and the
implication of the 2008 results on the VHE emission models are presented in
detail in \cite{Acc09}.

\begin{figure*}
\includegraphics[width=.85\linewidth]{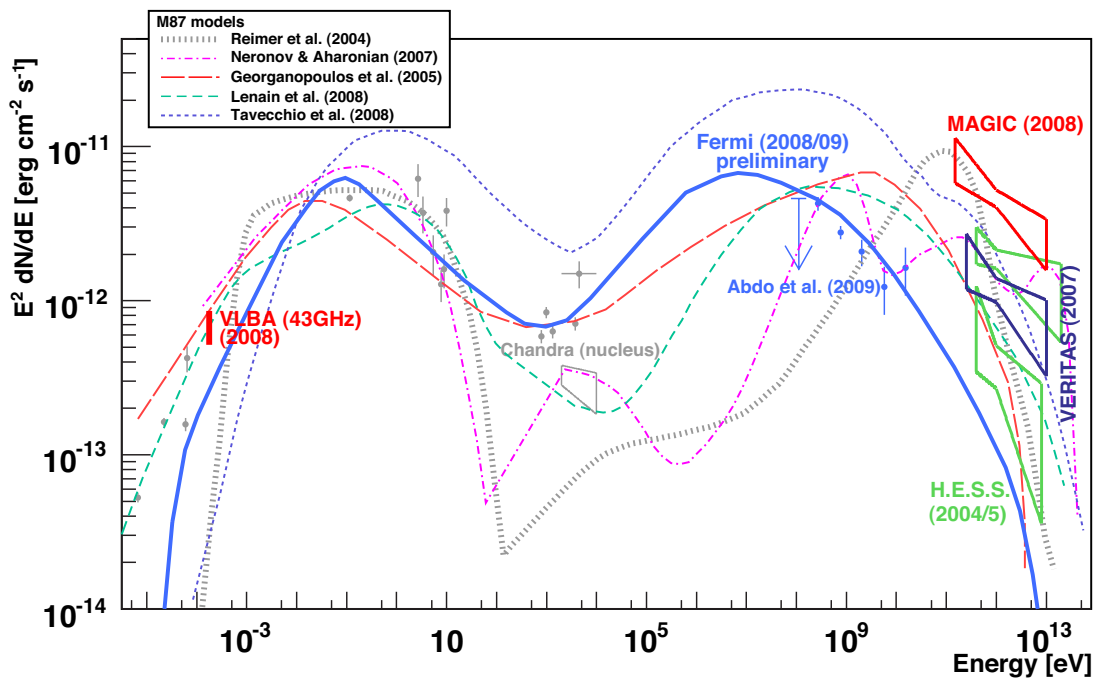}
\caption{
SED data and models for M 87. Grey points: archival radio and optical data, as well as the
EGRET upper limit, taken from \cite{Len08}. The vertical red line (VLBA) shows the range of radio fluxes during the 2008 flare (nucleus). Fermi data points (blue data points) and model (blue solid line) taken from \cite{F87}. The models are taken from the literature, see main text for references.
\label{fig:SEDs}}
\end{figure*}

\section{COLLECTION OF MODELS FOR THE GAMMA-RAY EMISSION IN M87}

The data collected in Fig. \ref{fig:SEDs} are non-simultaneous and the models
were partly published
before the TeV flaring was a known characteristics of M87. (Some of them
can probably be tuned to descibe the recent data). However, some
qualitative statements are possible in the light of the results of the
2008 campaign, motivating future
simultaneous observations. 

The hadronic synchrotron proton blazar model by Reimer et al. \cite{Rei04} 
was fit only based on the HEGRA detection (1998/99). However, it
seems to be hard to make it work in the Fermi energy range (very
different slopes) and it also leads to a strong intrinsic cut-off at a
few TeV, making it difficult to describe the hard TeV spectra
consistently measured by H.E.S.S., MAGIC and VERITAS. It seems to be
difficult to describe a radio/TeV connection in this model framework.

In the black hole magnetosphere model by Neronov \& Aharonian \cite{Ner07}, the TeV emission 
originates direct in BH vicinity. The data was fit to
the H.E.S.S. 2005 flare. The fit seems to be compatible with the general
shape of the Fermi regime, and it also is in the same ballpark for the
radio data.

The leptonic de-accelerated inner jet model by Georganopoulos, Perlman, \& Kazanas \cite{Geo05} was published 
before the H.E.S.S. 2005 flare, so it does not describe the hard TeV spectra well (strong cut-off). It might be
tuned, however, to a harder flare state.

In the model by Lenain et al. \cite{Len08}, individual blobs in the inner jet close to the
BH emit the TeV $\gamma$-rays. This model was fit to the H.E.S.S. low (2004) and high (2005)
states. Only the 2004 fit (low state) is currently shown in the plot in
order to declutter the plot. This model also explains the observed radio emission.

In the spine/sheath model \cite{Tav08} framework (Tavecchio \& Ghisellini)
it seems to be difficult to 
explain the radio data as part of this emission mechanism.

\section{CONCLUSION AND OUTLOOK}
The cooperation between H.E.S.S., MAGIC and VERITAS in 2008 allowed for an
optimized observation strategy, which resulted in the detection and detailed
measurement of a VHE $\gamma$-ray outburst from M\,87.
Simultaneous Chandra observations, found HST-1, the innermost knot in the jet,
in a low state, while the nucleus showed increased X-ray activity. This is in
contrast to the 2005 VHE $\gamma$-ray flare, where HST-1 was in an extreme high
state.  The radio activity in 2007--8, resolving the inner region of M87
down to some 10s Schwarzschild radii ($R_S$), allowed to infer the origin
of the VHE $\gamma$-ray emission. 
A model suggesting HST-1 to be the origin of the $\gamma$-ray emission seems
less likely in the light of the 2008 result.  Due to its proximity and the
viewing angle of the jet, M\,87 is a unique laboratory for studying the
connection between jet physics and the measured VHE $\gamma$-ray emission.
The modest localization accuracy of the TeV
instruments of many arcseconds, even for strong sources, is inadequate to
constrain the origin of the emission in the inner jets of AGNs. The
VLBA resolution
instead corresponds to 30 by 60 $R_S$. This is starting to
resolve the jet collimation region. The temporal coincidence of the TeV and
radio flares indicates that they are related and provides the first direct
evidence that the TeV radiation from this source is produced within a few tens
of $R_S$ of the radio core, thought to be coincident to within the VLBA resolution
with the black hole.



\section*{ACKNOWLEDGMENTS}
{\it H.E.S.S.:} The support of the Namibian authorities and of the
University of Namibia in facilitating the construction and operation of
H.E.S.S. is gratefully acknowledged, as is the support by the German
Ministry for Education and Research (BMBF), the Max Planck Society, the
French Ministry for Research, the CNRS-IN2P3 and the Astroparticle
Interdisciplinary Programme of the CNRS, the U.K.  Science and
Technology Facilities Council (STFC), the IPNP of the Charles
University, the Polish Ministry of Science and Higher Education, the
South African Department of Science and Technology and National Research
Foundation, and by the University of Namibia. We appreciate the
excellent work of the technical support staff in Berlin, Durham,
Hamburg, Heidelberg, Palaiseau, Paris, Saclay, and in Namibia in the
construction and operation of the equipment. \\
{\it MAGIC:} The collaboration acknowledges the excellent working conditions a the
Instituto de Astrofisica de Canarias' Observatorio del Roque de los Muchachos
in La Palma. The support of the German BMBF and MPG, the Italian INFN and
Spanish MCINN is gratefully acknowledged. This work was also supported by ETH
Research Grant TH 34/043, by the Polish MNiSzW Grant N~N203~390834, and by the
YIP of the Helmholtz Gemeinschaft. \\
{\it VERITAS:} 
This research is supported by grants from the US Department of Energy, the US
National Science Foundation, and the Smithsonian Institution, by NSERC in
Canada, by Science Foundation Ireland, and by STFC in the UK. We acknowledge
the excellent work of the technical support staff at the FLWO and the
collaborating institutions in the construction and operation of the instrument. \\
{\it VLBA team:} The Very Long Baseline Array is operated by The National Radio
Astronomy Observatory, which is a facility of the National Science Foundation
operated under cooperative agreement by Associated Universities, Inc.

R.M.W. acknowledges partial support by the DFG Cluster of Excellence
``Origin and Structure of the Universe''.

\end{document}